\documentclass[prb,twocolumn,showpacs,preprintnumbers,amsmath,amssymb,floatfix,superscriptaddress]{revtex4}

\usepackage[]{graphicx}         	
\usepackage{bm}             		
\usepackage{tabularx}
\usepackage{times}				

\newcommand{\IEF}{Institut d'Electronique Fondamentale, CNRS, UMR 8622, 91405 Orsay, France}
\newcommand{\UPS}{Universit{\'e} Paris-Sud, UMR 8622, 91405 Orsay, France}
\newcommand{\IMEC}{IMEC, Kapeldreef 75, B-3001 Leuven, Belgium}
\newcommand{\KULeuven}{Laboratorium voor Vaste-Stoffysica en Magnetisme, Katholieke Universiteit Leuven, B-3001 Leuven, Belgium}
\newcommand{\USFD}{Department of Engineering Materials, University of Sheffield, Sheffield S1 4DU, United Kingdom}

\begin{document}

%
%
\title{Current-driven vortex oscillations in metallic nanocontacts: Zero-field oscillations and training effects}

\author{M. van Kampen}
\affiliation{\IMEC}
\author{L. Lagae}
\affiliation{\IMEC}
\affiliation{\KULeuven}
\author{G. Hrkac}
\author{T. Schrefl}
\affiliation{\USFD}
\author{Joo-Von Kim}
\email{joo-von.kim@u-psud.fr}
\author{T. Devolder}
\author{C. Chappert}
\affiliation{\IEF}
\affiliation{\UPS}

\date{\today}

%
%
\begin{abstract}
We present an experimental and theoretical study of the low-field dynamics of current-driven vortex oscillations in nanocontacts based on spin-valve multilayers. These oscillations appear as low-frequency (250-500 MHz) excitations in the electrical power spectrum which arise from to variations in the giant-magnetoresistance. We show that the vortex oscillations, once nucleated at large fields applied perpendicular to the film plane, persist at zero applied magnetic fields. Some training effects on the oscillation frequency and linewidth also observed for small in-plane magnetic fields.
\end{abstract}


\maketitle

\section{Introduction}
Magnetic vortices are topological micromagnetic structures that arise from the competition between the exchange interaction and a planar anisotropy. In patterned magnetic materials, such as a magnetic dot for example, the stray dipolar fields at the sample edges defines a shape anisotropy that can lead to the formation of vortices. While magnetic vortices have been much studied over many decades, vast improvements in fabrication and lithography techniques in recent years have allowed for more detailed and controllable experimental studies, thereby generating a renewed interest in these topological objects. 

In patterned materials the dynamics of vortices is governed by the geometry of the sample, because the potential energy of the vortex is mainly determined by the dipolar stray field energy of the system. In addition to spin-wave modes, vortex excitations also include a translational motion whereby the vortex core precesses about its equilibrium position. In submicronic structures, it has been shown in a number of experiments and theoretical works that the confining potential gives rise to vortex oscillations, which involve translational motion, in the range of $0.1-1$ GHz\cite{Guslienko:JAP:2002,Novosad:PRB:2002b,Park:PRB:2003,Ivanov:JAP:2004,Buchanan:NatPhys:2005,Novosad:PRB:2005,Guslienko:PRL:2006,Buchanan:PRB:2006}. These modes are typically lower than the frequency of the ferromagnetic resonance mode in such structures.

In continuous systems such as thin films, there is no confining potential (aside from possible material defects) for a vortex structure, \emph{a priori}, since the translational symmetry in the film plane assures that the vortex energy is independent of its lateral position in the film. This is not the case for the nanocontact system, however, in which large electric currents are injected through a metallic point, typically with radii of 40-100 nm in size, in contact with the magnetic film~\cite{Rippard:PRL:2004}. Because a magnetoresistive spin-valve stack is usually employed in nanocontact systems, the electric current becomes spin-polarised upon its passage through the pinned magnetic layer and transfers spin-angular momentum to the free magnetic layer as it impinges upon this layer, through what is referred to as the ``spin-transfer'' effect~\cite{Slonczewski:JMMM:1996, Berger:PRB:1996}. The large current densities required for observable spin-transfer effects also mean that large Oersted-Amp{\`e}re fields are generated by the current, which break the translational symmetry of the film and lead to a magnetic potential for a vortex through the Zeeman interaction. For a nanocontact with a radius of 100 nm, for example, the Oersted-Amp{\`e}re field at the contact edges, assuming the contact to be an infinite circular wire, is approximately 60 mT for an applied current of 30 mA. For soft magnetic materials such as permalloy, such fields are much larger than the coercive fields required for magnetisation reversal.

The dynamics of magnetic vortices presents a number of interesting features from the point of view of fundamental and applied physics. Interesting questions concerning the fundamental aspects of single and pair-vortex dynamics, driven by magnetic fields~\cite{Buchanan:NatPhys:2005} or spin-polarised currents~\cite{Pribiag:NatPhys:2007}, have been brought to light in recent experiments. The latter example concerning current-driven dynamics is particularly interesting, because, to the best of our knowledge, it pertains to the first experimental demonstration of self-oscillatory dynamics of a magnetic vortex~\cite{Pribiag:NatPhys:2007}. The spin-transfer effect makes such self-oscillatory dynamics possible, because the additional torques introduced by the spin-transfer can be made to compensate the intrinsic damping associated with magnetisation dynamics under certain conditions. In light of these results, a number of applications for nanoscale oscillators and resonators based on magnetic vortices are envisaged.

The present authors have shown that the combination of Oersted-Amp{\`e}re fields and spin-transfer torques in metallic nanocontact systems can give rise to controllable vortex dynamics in a laterally extended system~\cite{Mistral:PRL:2008}. In that experiment, the spin-transfer torque is applied only locally within the nanocontact region, while the Oersted-Amp{\`e}re field generated by the applied electric current leads to a long-range magnetic potential in which well-defined vortex oscillations can be established. Furthermore, it was shown that the dynamics involve a large-amplitude translation of the vortex core along an orbit outside the nanocontact itself, leading to vortex oscillations with frequencies in the range of 250-500 MHz. Similar low frequency oscillations have been reported in a different experiment~\cite{Pufall:PRB:2007}. In this article, we present an experimental and numerical study of the low-field dynamics of such current-driven vortex motion in metallic nanocontact systems. We present evidence of a threshold for vortex nucleation and show that once nucleated the vortex oscillations persist at zero applied fields. We study these oscillations at small in-plane fields and present evidence of training effects. It is shown that the experimental results are consistent with micromagnetics simulations, in which a clear vortex orbit is shown to be established by the spin-transfer torques. Moreover, the simulations highlight the importance of the details concerning the spin-transfer torques, whereby a comparison between different mathematical forms of the spin-transfer are shown to lead to differences in orbital behaviour.

\section{Samples and experimental setup}
The experimental system studied comprises a metallic nanocontact deposited on a metallic spin-valve stack. The multilayer is composed of Ta(3.5 nm) / Cu(40) / Ta(3.5) / Ni$_{80}$Fe$_{20}$(3) / IrMn(6) / Co$_{90}$Fe$_{10}$(4.5) / Cu(3.5) / Co$_{90}$Fe$_{10}$(1.5) / Ni$_{80}$Fe$_{20}$(2) / Pt(3), where the figures in parentheses denote the layer thicknesses in nanometres. The composite CoFe/NiFe layer is the magnetic free layer of the spin-valve, while the 4.5 nm thick CoFe layer, which is exchange biased by the IrMn antiferromagnet, serves as a reference layer for the giant magnetoresistance variations and spin-transfer torques. The films are sputter-grown in an ultrahigh vacuum system with a base pressure of $3 \times 10^{-8}$~Torr. The sample mesa is rectangular and is $27 \times 17$ $\mu \textrm{m}$ in size. The contacts are made to the laterally extended film stack which is incorporated in the central conductor of a coplanar waveguide (CPW) structure. After patterning by conventional lift-off lithography, the stack is passivated by a 50 nm SiO$_2$ layer deposited using rf sputtering. The point contacts are defined in a poly(methyl methacrylate) (PMMA) layer by electron-beam lithography and etched into the SiO$_2$ by a short dip in a buffered HF solution. The coplanar waveguide structure is then deposited and patterned by lift-off lithography. The devices are annealed at 250$^{\circ}$ C for 10 minutes to improve the exchange bias coupling between the IrMn and the Co$_{90}$Fe$_{10}$ layers. 

Studies of the point-contact morphology in our systems have been made in detail using scanning electron microscopy, atomic force microscopy (AFM), ellipsometry and optical microscopy measurements. The target diameter of the point contacts are nominally 100 nm, however, due to the isotropic wet etch the diameter of the contacts is increased with respect to the 100 nm holes in the PMMA layer. Indeed, a significant variation in the diameter as a function of depth is found, with AFM measurements indicating a typical variation between 130 to 270 nm over a thickness of 40 nm. Based on the measured depth profile and the multilayer structure, the contact diameter at the free magnetic layer is estimated to be approximately 160 nm.

The magnetic properties of the free and reference layers were characterised using standard magnetometry methods. From hard-axis hysteresis loop measurements, we estimate the uniaxial anisotropy field for the free layer to be in the range $ 0 < \mu_0 H_{k} \leq $ 0.7 mT, while easy-axis loop measurements reveal a small bias field of 1.7 mT acting on the free layer, which we attribute to N\'eel coupling between the free and fixed layers. The exchange bias field acting on the reference layer is found to be $-60$ mT. The free layer saturation magnetisation is found to be $\mu_0 M_s = 1.5$ T from voltage noise spectroscopy measurements. 

For the electrical measurements, the current is applied perpendicular to the spin-valve stack  and is generated by applying a DC voltage to a $200~\Omega$ resistor in series with the sample. Two bias tees allow for DC and RF routing. Along one of the RF routes, an amplifier bank is used to amplify signals by +43~dB over the frequency range of 0.1-26 GHz, which is subsequently connected to a spectrum analyser. The other RF route is terminated by a 50 $\Omega$ load while the DC route is shorted. The final power spectra are obtained after subtracting a reference curve taken with a zero current bias. The RF measurements are made under applied magnetic fields in the film plane and perpendicular to the film plane. A 20~m$\Omega$ magnetoresistance is typical for our spin-valves. In our convention, positive currents describe the flow of electrons from the fixed to the free layer.

\section{High-frequency measurements}
As described elsewhere~\cite{Mistral:PRL:2008}, the vortex oscillations we study are characterised by sub-gigahertz excitations in the electrical power spectrum of the multilayer structure. The excitation mode consists of a translational motion whereby the vortex orbits the nanocontact outside the contact region. The Zeeman energy from the Oersted-Amp{\`e}re field, generated by the applied electrical current, gives rise to a central potential for the vortex motion. The addition of the spin-transfer effect gives rise to a torque that leads to a steady-state orbital motion of the vortex. The change in magnetisation direction underneath the contact, as a result of this vortex motion, gives rise to a time-varying giant-magnetoresistance with appears as high-frequency voltage oscillations. The power spectrum of these oscillations is studied with a spectrum analyser.

\begin{figure}
\centering\includegraphics[width=8.5cm]{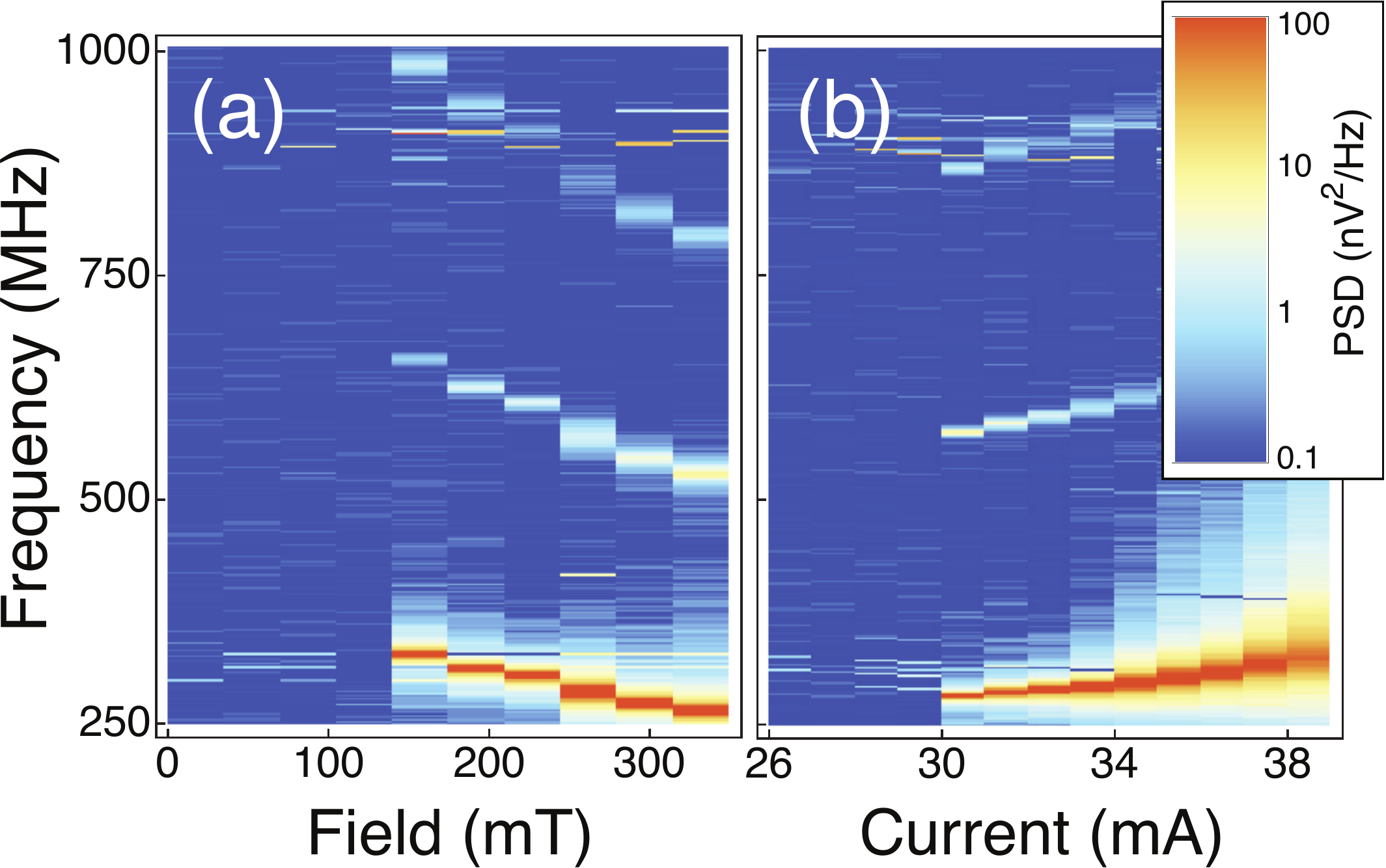}
\caption{\label{fig:spectra_map}(Colour online) Power spectral density (PSD) as a function of (a) perpendicular applied field, at a fixed current of 31 mA, and (b) applied current, at a fixed applied perpendicular field of 210 mT.}
\end{figure}

An example of the low-frequency power spectra that we observe is given in Fig.~\ref{fig:spectra_map}. 
%
The oscillations are typically in the range of 250-500 MHz, with linewidths of the order of 1 MHz, for the range of perpendicular fields and currents studied. Harmonics up to third order are detectable, but these are typically a few orders of magnitude smaller than the fundamental mode, which suggests that the magnetisation precession is largely circular. We would also like to stress that in most cases no other excitation modes are seen up to 1 GHz, so in the remainder of the paper the discussion will be focused entirely on this one excitation mode. In Fig.~\ref{fig:spectra_map}a, we present an example of a clear field threshold for oscillations at a given current (31 mA). Below a field of approximately 115 mT, the rf spectrum is quiet with no evidence of any oscillations. As the perpendicular field is increased past this threshold, we observe a large signal in the power spectrum, with a well-defined Lorentzian peak centred at around 300 MHz. This field threshold appears to be a ``hard'' threshold, in the sense that the excitation has a finite amplitude just above the threshold that does not increase gradually from the noise floor. In terms of the vortex model, we can understand this behaviour in terms of a ``nucleation'' field for the pseudo-vortex structure; below this field, it is energetically unfavourable for a pseudo-vortex to form. A similar hard threshold is seen for the current, as shown in Fig.~\ref{fig:spectra_map}b.

\begin{figure}
\centering\includegraphics[width=6.5cm]{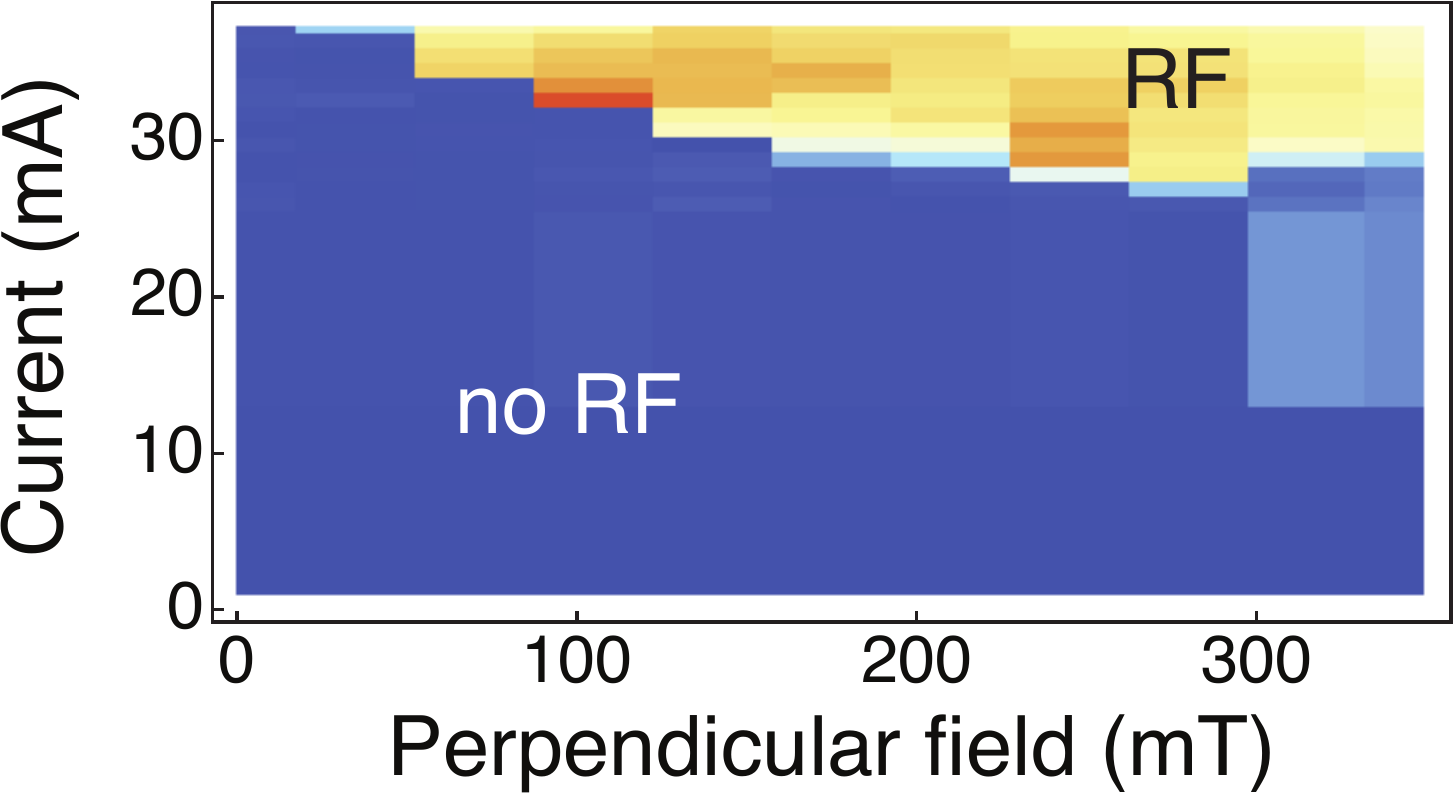}
\caption{\label{fig:HvI_map}(Colour online) Field-current map of integrated power in the high-frequency domain ($>$ 100 MHz). The region in which clear high-frequency (250 MHz - 1 GHz) excitations are observed is denoted by ``RF''.}
\end{figure}

This threshold behaviour is better illustrated in Fig.~\ref{fig:HvI_map}, where we present the perpendicular field-current phase diagram for the vortex oscillations. 
%
In this figure, the integrated power in the high-frequency region ($>$100 MHz) is shown as a colour-code as a function of applied perpendicular field and current. As discussed in the previous paragraph, the separation between the regions with and without high-frequency (rf) excitations appears to be quite sharp, which is suggestive of a nucleation process that is required before the vortex oscillations can take place.

\begin{figure}
\centering\includegraphics[width=7cm]{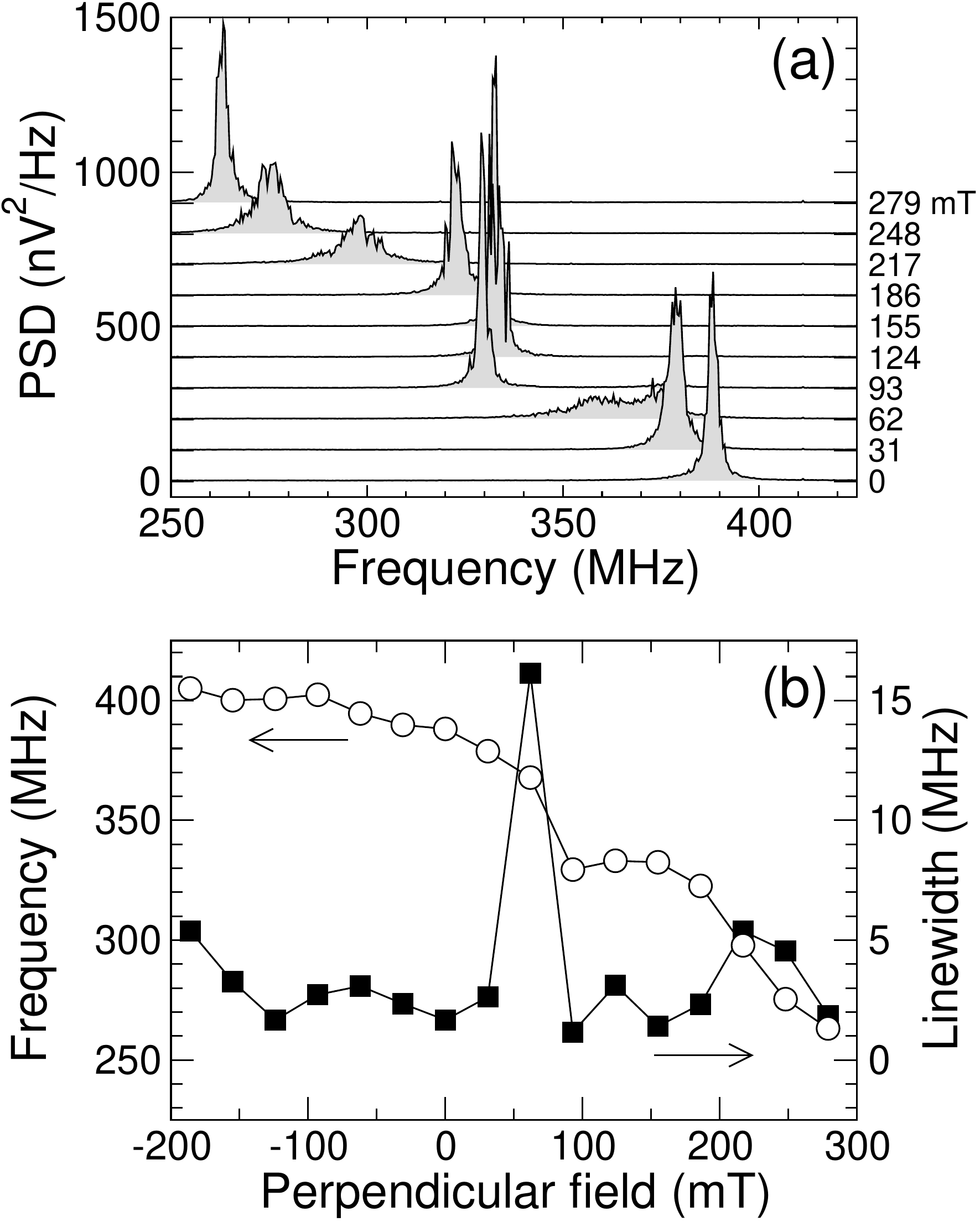}
\caption{\label{fig:psd_fieldvariation}(a) Power spectral density (PSD) for decreasing perpendicular applied fields under an applied current of 28 mA. A vertical offset has been applied to the curves for visual clarity. (b) Frequency and linewidth of the spectral lines in (a), obtained from Lorentzian fits.}
\end{figure}

Once nucleated at high fields, the vortex oscillations persist even at \emph{zero} applied field. An example of this behaviour is given in Fig.~\ref{fig:psd_fieldvariation}.
%
In this experiment, the vortex oscillation is nucleated for increasing fields up to 350 mT, and the oscillations are studied for gradually decreasing fields down to zero field. While there is a general monotonous increase in the oscillation frequency as the field is decreased, the variation is punctuated by a number of jumps in the frequency, as can be seen in Fig.~\ref{fig:psd_fieldvariation}a. This behaviour is suggestive of either a change in the micromagnetic structure of the vortex as the field is varied, or a change in the constraints that govern the vortex orbit around the point contact. This point is further highlighted in Fig.~\ref{fig:psd_fieldvariation}b in which we show the variation of the oscillation frequency and linewidth as a function of applied field. One observes a clear peak in the linewidth at the point at which a jump in frequency is observed. A monotonous increase in the frequency is observed as the field is reversed as expected from simple analytical theory~\cite{Mistral:PRL:2008}, however, the oscillation is only stable up to about 200 mT in reversed field. The fact that the frequency levels out at reversed field is also consistent with the vortex picture, as the field is opposite to the polarisation of the vortex. As such, one does not expect the frequency to depend much on the reversed field strength.

\begin{figure}
\centering\includegraphics[width=8.5cm]{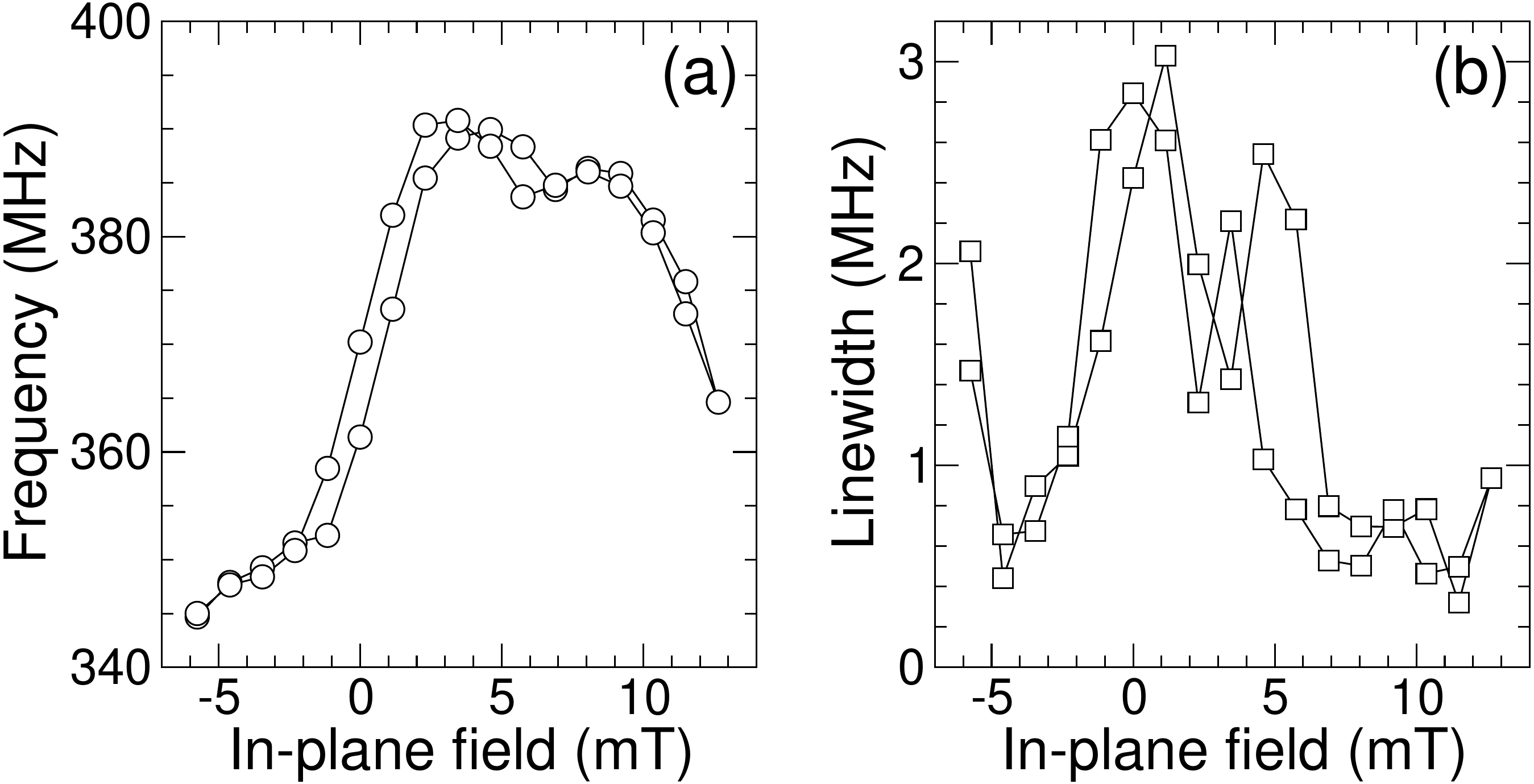}
\caption{\label{fig:inplane_sweep}Frequency and linewidth as a function of in-plane field sweep, for an applied current of 28 mA and zero perpendicular field.}
\end{figure}

With the low-frequency oscillation nucleated at large perpendicular fields, and then brought to zero field, we examined the in-plane field dependence of this excitation by cycling the in-plane field between -12.6 and 12.6 mT along the easy axis direction. The results are presented in Fig.~\ref{fig:inplane_sweep}.
%
A slight hysteresis in the frequency and linewidth of the oscillation mode is observed. In contrast to the case of uniform precession, for which one would expect a frequency variation at low applied in-plane fields $H_{||}$ of the form,
\begin{equation}
\frac{\partial \omega}{\partial H_{||}} \approx \frac{\gamma}{2}\sqrt{\frac{M_s}{H_k}},
\end{equation}
where $\gamma$ is the gyromagnetic ratio, which with our material parameters gives a slope of approximated 392 MHz/mT, which is clearly much larger than the frequency variation observed in experiment. 

\begin{figure}
\centering\includegraphics[width=7cm]{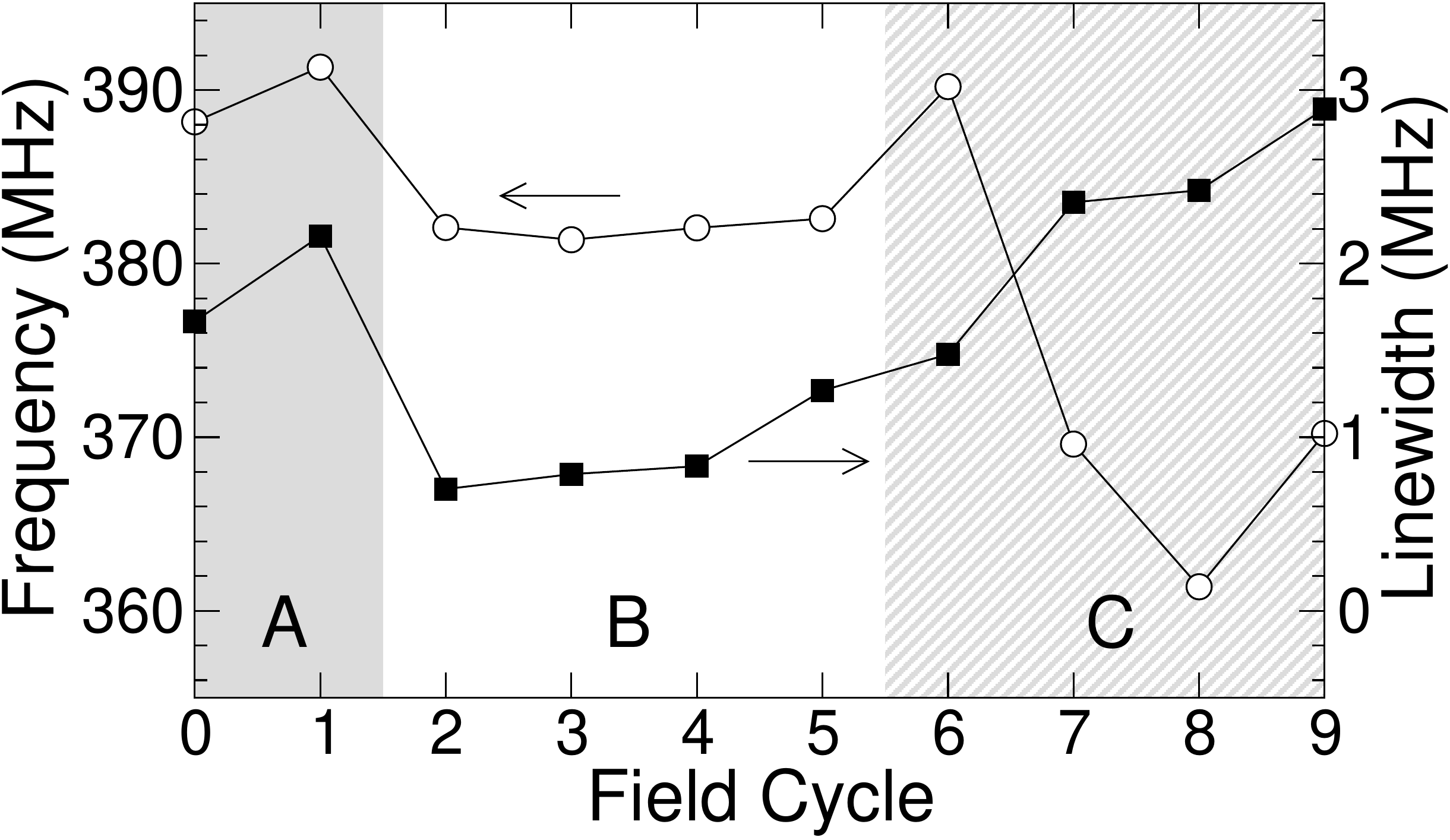}
\caption{\label{fig:training} (a) Oscillation frequency and (b) linewidth of main peak in power spectrum measured at zero applied field, for a number of different field histories.}
\end{figure}

With repeated cycling of the in-plane field, we observe some training effects on the frequency and linewidth of the oscillations. An example of this behaviour is shown in Fig.~\ref{fig:training}, where the oscillation frequency and linewidth are presented at zero applied fields, with different field histories. 
%
Three examples of different field histories are presented. In region A, the zero-field state is obtained after cycling the perpendicular applied field, first from large positive fields, through zero, then to large negative fields and back to zero. We observe a small difference in oscillation frequency and linewidth between the two cases, which suggests that a small change in micromagnetic structure may have taken place. In region B, the zero-field spectra are measured after repeated in-plane field cycles between $-$5 and 5 mT. In this region, in which the field was cycled twice, there is little variation in the oscillation frequency, with a small change in the linewidth. At these small cycling fields, we do not expect the vortex structure to lose stability, so there is no surprise that the frequencies remain approximately constant. In region C, the zero-field spectra are presented for larger in-plane field cycles, between $-$12  and 12 mT. At these applied fields, there is some evidence in the power spectra of a significant change in the micromagnetic structure, for example, the appearance of two peaks with similar frequencies. These double peaks only persist over a small range of applied fields (10-12 mT), and for field magnitudes smaller than these values, only one peak is observed. Nevertheless, the field cycling in this case presents more significant training effects, both in the frequency and the linewidth. Moreover, in contrast to regions A and B, the linewidth and frequency do not vary in the same manner.

\section{Micromagnetic simulations}
For a better understanding of the nature of low frequency oscillations found in experiments at zero field, we performed full micromagnetic simulations of a representative trilayer CoFe(3.5 nm)/Cu(3)/NiFe(4) stack. The inhomogeneous current distribution flowing through the free layer is computed self-consistently from diffusion equations and magnetoresistance variations. The Oersted-Amp{\`e}re field generated as the current flows through the contact is then computed self-consistently from this current distribution. Furthermore, the simulations were performed with the symmetric and asymmetric Slonczewski term for the spin-transfer~\cite{Xiao:PRB:2005,Ertl:JAP:2006}. The spin-transfer torque $(\partial \hat{m} / \partial t)_{\rm st} = \vec{N}_{\rm st}$ can be incorporated into the Landau-Lifshitz equation as an additional term of the form,
\begin{equation}
\vec{N}_{\rm st} = \eta(\theta) \frac{\hbar}{2e} \frac{J}{d} \hat{m} \times (\hat{m} \times \hat{p}),
\label{eq:STT}
\end{equation}
where $\hat{m}$ and $\hat{p}$ are unit vectors representing the orientation of the free and fixed layer magnetisations, respectively, $J$ is the current density, $d$ is the free layer film thickness, and $\theta$ is the angle between $\hat{m}$ and $\hat{p}$, $\cos\theta = \hat{m} \cdot \hat{p}$. For the symmetric Slonczewski torque, the function $\eta(\theta)$ is found to be
\begin{equation}
\eta_{\rm S}(\theta) = \frac{q}{A + B \cos\theta},
\end{equation}
while for the asymmetric torque, the angular dependence of the spin-transfer is
\begin{equation}
\eta_{\rm AS}(\theta) = \frac{q_+}{A + B \cos\theta} + \frac{q_-}{A - B\cos\theta},
\end{equation}
where $q$, $A$, and $B$ are material parameters that depend on the exact configuration of the multilayer structure~\cite{Xiao:PRB:2005}. The simulation area is a circular film 1000 nm in diameter, with the 160 nm diameter point contact at the centre. The system is discretised with a finite element method, using a linear basis function and a discretisation size of 4 nm, which is below the exchange length of 4.5 nm. An exchange-bias field of 162 mT is applied to the reference CoFe layer to simulate pinning in the real stacks. The voltage variations due to the GMR effect is computed by summing over all local variations in magnetisation between the free and reference layers. 

The simulations are performed as follows. First, we calculate the remanent state with an external field applied perpendicular to the film plane and in the absence of currents. From an initial uniform state in the film plane, we allow the magnetisation to relax toward the applied field direction through time-integration of the equations of motion. During this process, a vortex first forms under the point contact (which is situated in the centre of the simulation grid) and is followed by the creation and annihilation of multiple vortices. After this transient phase, a single vortex state emerges as the system relaxes to its remanent state. Next, a current of 30 mA is applied through the point contact with the remanent state serving as the initial state for this step of the calculation. We observe that the additional spin torque drives the vortex out of the contact area and towards a stable orbit around the contact. Then the external field is reduced to zero and remains zero for the whole computational time of 120 ns. These simulations show that the oscillations observed in experiment are related to a large-amplitude translational motion of a magnetic vortex, while a field is applied, shown in a previous paper~\cite{Mistral:PRL:2008}. Unlike the vortex dynamics observed in nanopillars where the vortex core precesses within the confining part of the Oersted field, the dynamics here correspond to an orbital motion outside the contact region.

\begin{figure}
\centering\includegraphics[width=7cm]{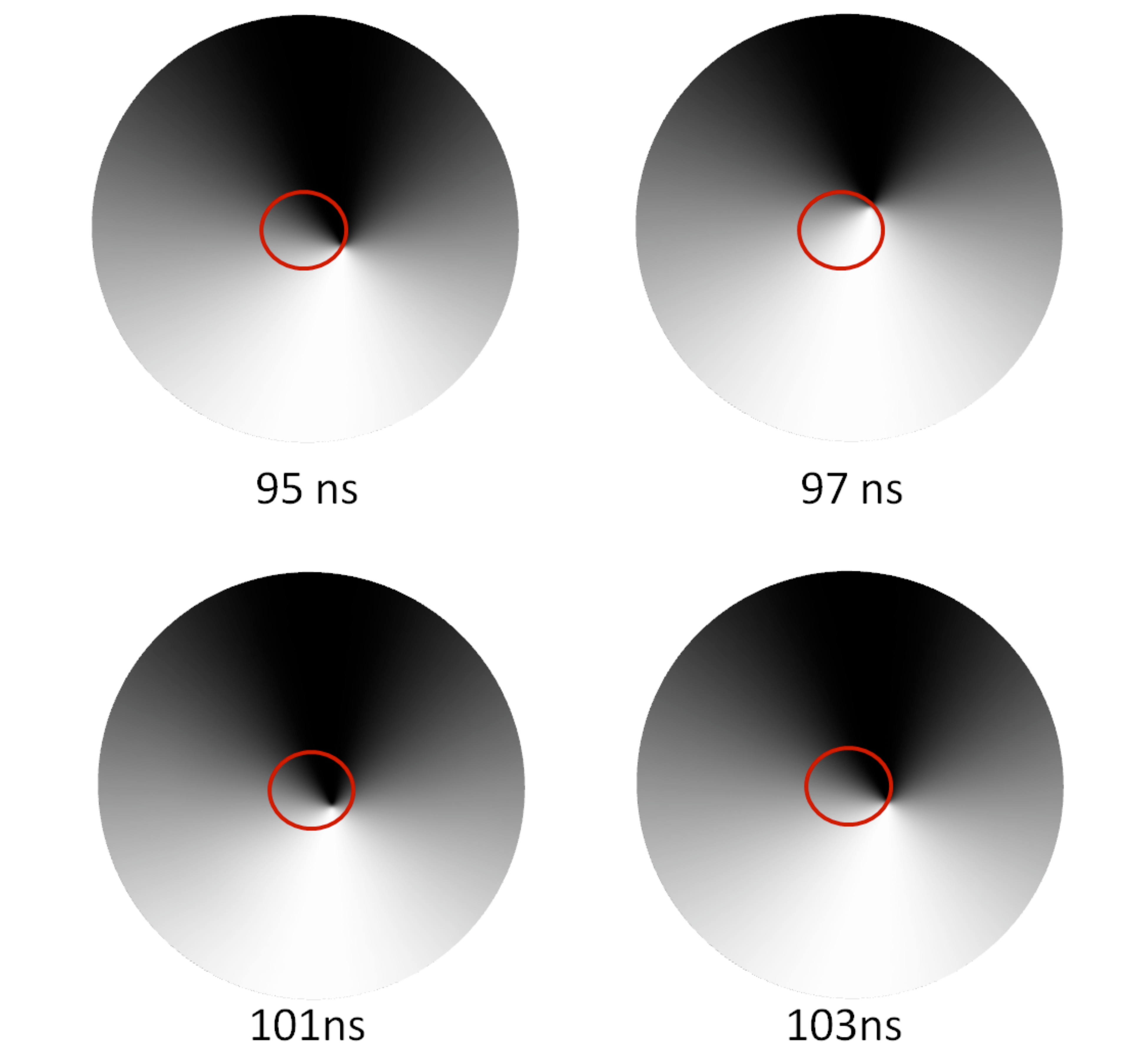}
\caption{\label{fig:vortex_orbit}(Colour online) Simulation results of vortex orbit at edge of point contact with an asymmetric Slonczewski torque. The grey scale indicates the magnetisation orientation in the film plane, with black representing magnetisation oriented to left, and white magnetisation oriented to the right. The red circle denotes the edges of the circular point contact.}
\end{figure}

\begin{figure}
\centering\includegraphics[width=8.5cm]{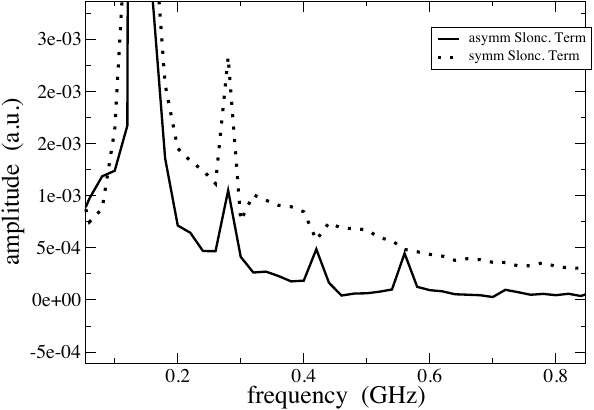}
\caption{\label{fig:simulation_psd} Power spectral density calculated from simulation, with a comparison between the symmetric and asymmetric Slonczewski terms.}
\end{figure}

Once the external field is reduced to zero the orbit of the vortex approaches the periphery of the physical point contact. We observe two distinct behaviours which are related to the choice of the Slonczewski term used to model the spin-transfer. For the symmetric term, the vortex orbits for 50 ns outside the point contact region, after which it moves under the point contact area where it continues along a circular orbit around the point contact centre. For the asymmetric term, the vortex immediately approaches the physical point contact boundary and resumes an elliptical trajectory around the contact centre, leading to a motion that traverses the regions both inside and outside the point contact area, as shown in Fig.~\ref{fig:vortex_orbit}.
%
A Fourier analysis of the high-frequency magnetoresistance oscillations gives a central frequency of 280 MHz for both cases, but higher harmonics at 421 and 560 MHz are more visible for the asymmetric case, as shown in Fig.~\ref{fig:simulation_psd}.
%
By comparing these results with experiment, one can see that good quantitative agreement between the simulation and experimental frequencies is achieved.

\section{Discussion and concluding remarks}
We believe that the low-field dynamics presented here lend further support to the hypothesis of current-driven vortex oscillations in our nanocontact systems. The two key observations presented here, namely the persistence of oscillations in zero applied field and the weak field dependence of the frequency for small in-plane fields, cannot be explained with a uniform precession model. Moreover, the frequencies of the oscillations observed at much lower than those expected for spin-wave modes at the same applied fields.

There are some features observed in experiment, however, that are not immediately captured in the simple analytical model for vortex oscillations, based on a Thiele equation~\cite{Mistral:PRL:2008}. For example, the appearance of steps in the frequency versus perpendicular field curves is suggestive of a change in the micromagnetic structure of the vortex (e.g., in the range of 100-200 mT in Fig.~\ref{fig:psd_fieldvariation}b). The theory presented in Ref.~\cite{Mistral:PRL:2008}, on the other hand, led to an expression for the frequency that is varies continuously as a function of perpendicular applied field (see Eq. 4 in Ref.~\cite{Mistral:PRL:2008}). The results of micromagnetics simulations presented here, which show a change in the vortex orbit as the perpendicular field is varied, give a hint as to the origin of these steps. We speculate that these steps may be a result of this change in orbit, whereupon executing an orbit near the contact region the vortex undergoes some small change in its micromagnetic structure. We point out that this can be accounted for in the analytical model by carefully integrating the vortex profile across the point-contact region. Further analytical and numerical studies on this point are currently in progress.

Another example concerns zero-field oscillations. It was argued in Ref.~\cite{Mistral:PRL:2008} that a component of spin-torque $p_{\perp}$, perpendicular to the film plane, is required for self-oscillatory dynamics. This is naturally present in a non-zero perpendicular magnetic field, because the magnetisation of the fixed layer is tilted out of the film plane under these conditions, which gives rise to a small perpendicular spin-torque component. In the absence of applied fields, however, one would expect this component to vanish, because it is more favourable energetically for the fixed layer magnetisation to lie in the film plane. But the condition of nonzero $p_{\perp}$ for self-oscillations was derived for the simple case where $\eta(\theta) = 1$ in Eq.~\ref{eq:STT}. We argue that the spin-transfer may have a modified form to that presented in Eq.~\ref{eq:STT} in realistic multilayer stacks, whereby a nontrivial angular dependence, such as a ``wavy'' variation,~\cite{Boulle:NatPhys:2007} might exist to give a non-vanishing $p_{\perp}$ even when the fixed layer magnetisation is in the film plane. This could be verified experimentally by changing the layer thickness of the spin-valve structure, by choosing different ferromagnetic materials for the free and fixed layers, or both. Indeed, our micromagnetic simulations, which use a nontrivial form for $\eta(\theta)$, confirm the possibility of exciting vortex oscillations at zero field.

In summary, we have presented an experimental and numerical study of the low-field dynamics of current-driven vortex oscillations in magnetic nanocontacts. We have shown that the vortex oscillations, once nucleated at sufficiently large fields, applied perpendicular to the film plane, persist down to zero fields as these are reduced. Some training effects, notably on the mode frequency and linewidth, are observed for small field sweeps in the film plane. Micromagnetics simulations confirm the possibility of zero-field vortex oscillations, and also show the possibility of a change in vortex orbit whereby the orbit begins from outside the contact and falls within the contact region.

\begin{acknowledgments}
This work was supported by the European Communities programs IST STREP, under Contract No. IST-016939 TUNAMOS, and ``Structuring the ERA'', under Contract No. MRTN-CT-2006-035327 SPINSWITCH, and by the local government of R{\'e}gion Ile-de-France within the framework of C'Nano IdF.
\end{acknowledgments}

\bibliography{articles}

\end{document}